\documentstyle[preprint,prl,aps]{revtex}
\newcommand{\beq}{\begin{equation}}
\newcommand{\eeq}{\end{equation}}
\newcommand{\eqna}{\begin{eqnarray}}
\newcommand{\eqne}{\end{eqnarray}}
\newcommand{\dia}{\begin{displaymath}}
\newcommand{\die}{\end{displaymath}}
\newcommand{\eqnaa}{\begin{eqnarray*}}
\newcommand{\eqnae}{\end{eqnarray*}}
\begin{document}
\draft
\title{Parametric Excitation of Edge Waves in the Quantum Hall Effect}

\author{Dror Orgad and Shimon Levit}

\address{Department of Condensed Matter, The Weizmann Institute of Science,\\
Rehovot 76100, Israel}

\date{\today}

\maketitle

\begin{abstract}
We consider the process of parametric excitation of gapless edge modes in a Hall bar 
by an alternating current. We find that such a process can be realized provided both 
an inter-edge interaction and a constant current are present in the bar. The expected 
experimental signatures of this phenomenon are discussed.

\end{abstract}

\pacs{PACS numbers: 73.40.Hm, 72.15.Nj}

Parametric excitation of waves is a phenomenon shared by many nonlinear 
systems. Generation of spin waves (magnons) in magnetically ordered dielectrics by 
microwave fields and the excitation of standing surface waves in liquids by vertical 
vibrations are just two examples out of a wealth of such instances (see, e.g. 
\cite{Lvov94}). A prerequisite for the existence of parametric processes in a system is a 
dependence of the dispersion law of its excitations on some externally controllable parameter.
If such a condition holds a suitable temporal and spatial variation of this parameter 
may result in the generation of excitations. The instrumental role of edges in forming 
the gapless modes of both the integer and fractional quantum Hall effects 
(IQHE,FQHE) was noted theoretically a long time ago \cite{Halperin82,Wen91a}. 
Subsequently these, so called edge magnetoplasmons, were observed in a series of 
experiments \cite{Allen83,Govorkov86,Volkov86,Govorkov87,Ashoori92,Talyanskii92}. 
In classical \cite{Volkov88} and quantum \cite{Wen91b,Oreg95} treatments of the 
problem it was shown that inter-edge interaction in a Hall bar causes mixing of 
the single edge density waves. The resulting excitations remain gapless and have components 
on both edges propagating in the same direction.
In a recent paper we considered these gapless modes and noted that 
their dispersion relation depends on the total current that flows in the bar 
\cite{Orgad95}. This observation led us to suggest the possibility of parametrically 
exciting these edge waves by the application of an alternating current through the bar. 
It is the purpose of this letter to show how this may be realized and to discuss the 
experimental signatures that such a phenomenon is expected to exhibit. 

Although our previous work \cite{Orgad95} was carried out in the framework of the 
Chern-Simons theory we find it more convenient to explore the present subject 
using the Luttinger model description of the QHE. We will consider a Hall bar 
of width $W$ and length $L(\gg W)$ under the conditions of the IQHE or the FQHE. 
Furthermore, we restrict our  discussion to the case where only the first Landau 
level is filled or to FQHE plateaus with filling factors $\nu=1/n$ where 
$n$ is an odd integer. Under these conditions each edge is described by one 
branch of single particle states which is obtained by linearizing the energy of the 
Landau level for non interacting electrons around the Fermi energy at the edge \cite{Wen92}. 
Each of the linear edge branches (Left and Right) ends at a lower cutoff in energy 
corresponding to the points where a degenerate level of bulk states commences.

The Hamiltonian of the Luttinger model which describes the edge excitations in this case 
is given by \cite{Wen92,Stone92}  
\begin{eqnarray}
\label{hamilton}
\nonumber H_0 &&= \sum_{k>0}\;\sum_{i=L,R}\left[{2\pi\hbar\over{\nu}}v_i + V(k)\right]
                \rho_i(k)\rho_i(-k) \\  
            && + \sum_{k}U(k)\rho_L(k)\rho_R(-k) \;\; ,
\end{eqnarray}
where the operators $\rho_{L,R}(k)$ create chiral waves concentrated along one of the 
edges. The constants $v_{L,R}$ are the velocities of the two branches of the spectrum, 
$V(k)$ is the  k--dependent  part of the intra-edge interaction (see below) and $U(k)$ 
is the interaction between the edges. Previous studies have demonstrated that the 
contributions to the velocities $v_{L,R}$ originate from two different sources.
The first is the bare Fermi velocities $v_{L,R}^{(0)}$ of the non interacting branches.
These velocities may be different due to differences in the confining potentials or in 
the Fermi energies of the two edges. Since in the absence of interactions the total 
current flowing in the bar is given by $I=\sigma_{xy}(E_{F_R}-E_{F_L})/e$ it is clear 
that the latter cause persists when such a current is present.

The second contribution to $v_{L,R}$ comes from the renormalization of the edge potential 
by the $k$--independent component of the intra-edge interaction. In the present work each 
electron is assumed to interact via unscreened Coulomb interaction with the other electrons 
of its edge and with the electrons of the opposite edge. In our Chern-Simons study 
\cite{Orgad95} we found that small deviations from the static condensate density 
$\rho(x;x_L,x_R)$ of the the form $\delta\rho_{L,R}({\bf r};t) =\eta_{L,R}(x)\xi_{L,R}(y,t)$ 
correspond to the gapless edge modes. Here $x_{L,R}$ are the "collective coordinates" of 
the condensate which control the excess charge and current at the edges. 
The functions $\eta_{L,R}(x)=\partial\rho(x;x_L,x_R)/\partial x_{L,R}$ 
describe the profiles of the excitations in the direction perpendicular to the walls while 
$\xi_{L,R}(y,t)$ are the amplitudes of the waves along the edges. Accordingly one finds that 
the self interaction of these waves is given by $1/2\int dk\, \xi_{L,R}(k)[2\pi\hbar\nu^{-1}
v_{L,R}^{(1)}+V(k)]\xi_{L,R}(-k)$ where the effective one-dimensional intra-edge 
interaction is 
\begin{eqnarray}
\label{intrap}
\nonumber && v_{L,R}^{(1)}=-{{\nu e^2}\over{\pi\hbar\epsilon}} \int dx dx' \, \eta_{L,R}
               (x)\ln\left({{|x-x'|}\over l}\right)\eta_{L,R}(x') \; , \\
          && V(k)= {{2e^2}\over \epsilon}\ln\left({2e^{-\gamma}\over{|k|l}}\right) \;.
\end{eqnarray}
Eq. (\ref{intrap}) is valid in the long wave length limit $k\to 0$ and $\gamma$ is 
the Euler constant that arises in this limit. We denoted the dielectric constant by 
$\epsilon$ and introduced the magnetic length $l=\sqrt{\hbar/m\omega_c}$. The 
$k$--independent parts $v_{L,R}^{(1)}$ are the contributions of the intra-edge interaction 
to the renormalization of the edge velocities. As we showed in \cite{Orgad95} the condition of 
electrical neutrality and the value of the total current $I$ determine the values of $x_{L,R}$ 
and thereby also of $v_{L,R}^{(1)}$.   

For the inter-edge interaction one obtains, assuming $|x_{L,R}| \ll W$, 
\begin{equation}
\label{interp}
U(k)={{2e^2}\over{\epsilon}}K_0(|k|W) 
\end{equation}
where $K_0$ is the modified Bessel function.

We now wish to consider the changes in the model which must be introduced in order to discuss
the effect of a small alternating current component $\Delta I\sin\Omega t$ added to the 
constant current $I_0$. According to our discussion above such a current should cause the 
velocities $v_{L,R}=v_{L,R}^{(0)}+v_{L,R}^{(1)}$ to become time dependent. To first order in 
$\Delta I$ and for $\Omega\ll\omega_c$ this will amount to the replacement $v_{L,R}\to v_{L,R}+ 
\Delta v_{L,R}\sin\Omega t$ in the Hamiltonian (\ref{hamilton}), with $\Delta v_{L,R} = \Delta I\,
\partial v_{L,R} / \partial I$. We adopt the resulting time independent Hamiltonian as the 
starting point for our analysis and will now demonstrate that it leads to resonant parametric 
generation of gapless edge waves. Both the velocity variations $\Delta v_{L,R}$ and $I_0$ 
will play an essential role in determining the effective pumping strength for this process.   

 As a first step we  utilize a Bogoliubov transformation to a new set of operators 
\begin{equation}
\label{trans}
\rho_{1,2}(k)=\sinh\theta_k \rho_{L,R}(k) + \cosh\theta_k \rho_{R,L}(k) \; .
\end{equation}
The angles $\theta_k$ are chosen in a way that diagonalizes the time independent part of $H$
\begin{equation}
\label{theta}
\sinh 2\theta_k={U(k)\over{\sqrt{[\pi\hbar\nu^{-1}(v_L+v_R)+V(k)]^2-U^2(k)}}} \;\; .
\end{equation}
The new modes, described by $\rho_{1,2}$, can still be classified as left and right movers. 
In contrast to the excitations created by $\rho_{L,R}$, each of these modes is composed 
of two components that live simultaneously on opposite edges while propagating in the 
same direction, Refs. \cite{Volkov88,Wen91b,Oreg95,Orgad95}. The velocities of the new modes 
are given by
\begin{eqnarray}
\label{newv}
v_{1,2}(k)=&& \pm{1\over 2}(v_R-v_L) \\
\nonumber && +{\nu\over{2\pi\hbar}}\sqrt{\left[{\pi\hbar\over \nu}(v_L+v_R)+V(k)\right]^2
             -U^2(k)} \;\;\; .
\end{eqnarray}
It is convenient to rescale the density operators in order to make them obey 
the familiar bosonic commutation relations. Thus we set, for $k>0$, 
          $\rho_1(k)=(k\nu/2\pi)^{1/2}b_{1,k}^{\dagger}$  and  
          $\rho_2(k)=(k\nu/2\pi)^{1/2}b_{2,-k}$ 
with $\rho_{1,2}(-k)=\rho_{1,2}^{\dagger}(k)$. Finally by introducing the definitions
\begin{eqnarray}
\label{defs}
\nonumber && \omega_{1,2}(k)=k v_{1,2}(k) \;\; , \;\;
             \Delta(k)=-{k\over 2}(\Delta v_L+\Delta v_R)\sinh 2\theta_k \; ,\\
          && \Delta_{1,2}(k)=k(\Delta v_{L,R}\sinh^2\theta_k+\Delta v_{R,L}\cosh^2\theta_k)
\end{eqnarray}
we obtain the following expression for $H$ which is best suited for our application
\begin{eqnarray}
\label{finalH}
\nonumber H= \hbar\sum_{k>0} && \{[\omega_1(k)+\Delta_1(k)\sin\Omega t]\, b_{1,k}^{\dagger} 
              b_{1,k}\\ 
\nonumber  && +[\omega_2(k)+\Delta_2(k)\sin\Omega t]\, b_{2,-k}^{\dagger}b_{2,-k} \\
           && +\Delta(k)\sin\Omega t \, (b_{1,k}^{\dagger}b_{2,-k}^{\dagger}+
              b_{2,-k}b_{1,k})\} \; .
\end{eqnarray}
In this form the effects of the alternating component of the current are apparent. 
The periodic alteration of the renormalized Fermi velocities induced by this component 
results in an oscillatory shift of the frequencies of the modes. The crux of the matter, 
though, resides in the last term of the Hamiltonian. This term describes the continuous 
parametric conversion of the oscillating pumping field into pairs of edge modes. Inspection 
of Eqs. (\ref{theta}) and (\ref{defs}) reveals that the effective pumping amplitude 
$\Delta(k)$ for such processes is determined by two ingredients. The first is the the strength 
of the inter-edge interaction $U(k)$. As this interaction is turned off the two edges 
decouple, $\Delta(k)$ vanishes and the only residual effect of the current is the 
periodic modulation of $\omega_{1,2}$. The absence of parametric excitation in this limit is 
a manifestation of momentum conservation and the chirality of the single-edge modes. 
One cannot convert a zero $k$ quantum of the perturbing field into a pair 
of waves while preserving the total momentum without the coexistence of modes with 
opposite momenta in the non separable interacting bar. 

The second ingredient necessary for the realization of the parametric excitation is the 
existence of asymmetry between the response of the two edges to small changes in the current.
For identical edges electrical neutrality implies $v_L(I)=v_R(-I)$, \cite{Orgad95}, 
which in turn means that $\partial v_L/\partial I=-\partial v_R/\partial I$ for the case
where $I=0$. As a result $\Delta(k)$ will vanish. One 
has to break the symmetry between the edges in order to obtain effective pumping. This 
statement is analogous to the well known result from the field of nonlinear optics that 
in crystals possessing an inversion symmetry (centrosymmetric) there is no parametric 
oscillation. The symmetry breaking can be achieved either through the use of samples with 
edges defined by different means (for example etched vs. gate-induced edges) or more easily 
by applying constant current $I_0$ in the bar.

Our next goal is to examine the time evolution of the parametric resonance and to identify 
its unique characteristics. To this end we consider the expectation values $n_{1,2}(k,t)$ 
of the number operators for the two modes and define a BCS-like anomalous correlator 
$\sigma(k,t)$
\begin{eqnarray}
\label{cordef}
\nonumber && n_1(k,t)=\langle b_{1,k}(t)b_{1,k}^{\dagger}(t)\rangle \; , \;
n_2(k,t)=\langle b_{2,-k}^{\dagger}(t)b_{2,-k}(t)\rangle \; , \\
          && \sigma(k,t)=\langle b_{1,k}(t)b_{2,-k}(t)\rangle 
\end{eqnarray}
where the average is taken with respect to the initial state of the system. The equations 
of motion for these quantities are readily derived using the Heisenberg equations for the 
$b_{1,2}$ operators. We simulate the presence of damping by adding imaginary parts 
$-i\Gamma_{1,2}(k)/2$ to the frequencies of the modes and assuming 
$\Gamma_{1,2}(k)=\Gamma(k)$. As a result we obtain
\begin{eqnarray}
\label{eom}
           \partial_t n_{1,2}(k,t)=&& -\Gamma(k)n_{1,2}(k,t)-2\Delta(k)\sin\Omega t\,
             {\rm Im}\sigma(k,t) \\
\nonumber  \partial_t {\rm Im}\sigma(k,t)= &&-\Gamma(k){\rm Im}\sigma(k,t)-[\omega_1(k)+
              \omega_2(k)+(\Delta_1(k)+\Delta_2(k))\sin\Omega t]{\rm Re}\sigma(k,t) \\
\nonumber  && -\Delta(k)\sin\Omega t[n_1(k,t)+n_2(k,t)] \\
\nonumber  \partial_t {\rm Re}\sigma(k,t)=&& -\Gamma(k){\rm Re}\sigma(k,t)+
              [\omega_1(k)+\omega_2(k)+(\Delta_1(k)+\Delta_2(k))\sin\Omega t]{\rm Im}
              \sigma(k,t) \; .
\end{eqnarray}
These equations are clearly decoupled for different values of $k$ thus allowing for 
independent consideration of different conversion processes. As is well known, parametric 
resonance is strongest when energy conservation is nearly fulfilled. Hence, concentrating on 
a specific $k$, we put $\Omega=\omega_1(k)+\omega_2(k)+\epsilon(k)$, where $\epsilon(k)\ll
\omega_{1,2}(k)$. We also require $\Delta(k),\Delta_{1,2}(k)\ll\omega_{1,2}(k)$.  
Up to lowest order in the small quantities one finds
\begin{eqnarray}
\label{sol}
          && \sigma(k,t)= e^{[s(k)-\Gamma(k)]t}\times \\
\nonumber && \hspace{1.8cm} \times\left[({\epsilon(k)\over{s(k)}}-i)\sin\Omega t+
               (1+{\epsilon(k)\over{s(k)}}i)\cos\Omega t\right] \\
\nonumber && n_{1,2}(k,t)={\Delta(k)\over{s(k)}}e^{[s(k)-\Gamma(k)]t} \;\; \;;
             \;\; s^2(k)=\Delta^2(k)-\epsilon^2(k) \; .
\end{eqnarray}
This solution is determined up to a multiplicative factor since the set (\ref{eom}) is 
linear. It is valid as long as $s(k)$ is real. For imaginary $s(k)$ one obtains an 
oscillatory decaying solution.  Going to higher orders in $s(k)$ will result 
in the appearance of other harmonics of $\Omega$ both in $n_{1,2}$ and in $\sigma$. 
The limit of the region of instability, where the solution (\ref{sol}) describes exponential 
growth of the amplitude of the edge waves, is given by the equation $s(k)-\Gamma(k)=0$. 
Thus for the resonance range in the frequency domain we have
\begin{equation}
\label{freqreg}
|\epsilon(k)|<\sqrt{\Delta^2(k)-\Gamma^2(k)} \; .
\end{equation}
It is a prominent feature of parametric excitation that dissipation does not eliminate the 
divergences near the resonance but only introduces a threshold  
\begin{equation}
\label{thresh}
\Delta_{th}(k)=\Gamma(k) 
\end{equation}
for the pumping amplitude below which instability does not develop. 
The exponential growth of $n_{1,2}$ in the resonance region reflects an ever increasing 
transfer of energy from the pumping field to the edge excitations. 
Non linear processes described by higher order terms in the Hamiltonian (\ref{hamilton})
will limit these divergences. This issue will not be addressed  in the present work.

The existence of a threshold and its dependence on external parameters, such as frequency  
and dc current, constitute the primary evidence for the presence of parametric resonance.
Since the threshold is determined by the amount of dissipation in the system a theoretical 
estimation of the latter is required. As we are unable to address this issue within the model  
used so far other approaches have to be considered. We choose to 
follow the work of Volkov and Mikhailov \cite{Volkov88} who investigated the edge 
magnetoplasmons using classical electrodynamics. In this scheme dissipation was incorporated 
via finite diagonal elements of the conductivity tensor. Using their results one 
recovers, for the case of zero current through the bar ($\omega_{1,2}=\omega$), the velocities 
(\ref{newv}) and finds
\begin{eqnarray}
\label{dissipation}
\nonumber   \Gamma(k)&\simeq&{2\over{\tau}}\ln^{-1}\left({2e\over{|k|l_0}}\right) 
            \;\;\; : \; \omega\tau\gg 1 \gg l_0\,{\rm max}(k,W^{-1}) \\
\nonumber   \Gamma(k)&\simeq&{{2\pi\sigma_{xy}|k|}\over{\epsilon}} \;\;\;\;\;\;\;\;\;\,
            \;\; : \; l_0\,{\rm max}(k,W^{-1})\ll \omega\tau\ll 1 \\
            &&
\end{eqnarray}
where $\tau$ reduces, in a strong magnetic field ($\omega_c\gg|\omega+i/\tau|$) and within 
the Drude model, to the elastic relaxation time. The length $l_0=e^2\nu/\epsilon
\hbar\omega_c$ determines the distance over which the electron-electron interaction is 
comparable to the cyclotron energy. Under the specified conditions $\tau=\nu[\omega_c 
{\rm Re}\,\sigma_{xx}/(e^2/h)]^{-1}$ and it is clear that the first region considered in 
(\ref{dissipation}) corresponds to the QHE plateaus while the second case applies to the 
transitions between the plateaus. We thus expect the threshold to follow the general 
behavior of $\sigma_{xx}$ as a function of the magnetic field. Inside the plateaus the 
threshold should scale with temperature in much the same way as $\sigma_{xx}$. Few 
theoretical and experimental studies have shown that the latter varies exponentially 
with temperature either due to thermal activation or as a result of variable range hopping 
\cite{Prange90}.

In order to calculate the contribution of $\Delta v_L +\Delta v_R$ to the pumping force 
one has to resort to microscopic calculations. We use numerical results from our previous 
work \cite{Orgad95}. The resulting pumping amplitude, for a narrow 
IQHE bar with hard walls, is presented in Fig. \ref{pumpgraph}. This quantity displays maxima 
both in $k$ and $I_0$. The maximum in $k$ is due to the competing effects of the explicit 
linear dependence of $\Delta$ on $k$ and the exponential decay of $U$ for $k\geq W^{-1}$. 
The maximum in $I_0$ reflects the fact that for high enough currents smaller changes in the 
parameters of the edges are needed in order to achieve the same change in $I$. These maxima 
translate into minima in the threshold for the amplitude of the alternating current above 
which parametric excitation is expected to take place. This is explicitly seen in Fig. 
\ref{thgraph} where we present the typical dependence of the threshold on the pumping 
frequency and the constant current through the bar. In the QHE plateaus and for the range 
of $k$ for which $\Delta$ is appreciable (cf. Fig. \ref{pumpgraph}), one finds that 
$\Gamma\approx 0.2\omega_c\,\sigma_{xx}/\sigma_{xy}$. Since in these regions, and for 
temperatures of a few degrees, the ratio $\sigma_{xx}/\sigma_{xy}$ varies in the range of 
$10^{-8}-10^{-4}$ \cite{Prange90} one expects to be able to excite the waves even though, 
due to the assumption of hard walls, the values of $\Delta$ shown in Fig. \ref{pumpgraph} 
are probably an overestimate for a realistic sample. 

Up to this point we considered a bar which is connected to a current source. Another way 
to induce the alternating current is by placing the bar in a microwave cavity.
The radiation then serves both as an external oscillating Hall field and, through its 
absorption by the sample, as a probe to the onset of parametric processes. 
The behavior, in this case, of $\Delta$ (per unit amplitude of the oscillating field) and 
of the threshold is similar to the one depicted in Figs. \ref{pumpgraph} and \ref{thgraph}. 
For this setup, however, we do not find a maximum of $\Delta$ in $I_0$ but instead 
observe a saturation of the pumping amplitude at large currents. 
We conclude by remarking that it may also be of interest to measure the 
dependence of the threshold on screening by varying the distance to a metallic back-gate. 

We are grateful to Y. Levinson, V. Cherepanov, V. Talyanskii, G. Falkovich, V. Steinberg 
and M. Reznikov for interesting and useful discussions.

\begin{figure}
\caption{The effective pumping amplitude $\Delta$ per unit amplitude of the oscillating 
current $\Delta I$ as a function of the wave vector $k$ of the excitations and the constant 
current $I_0$ flowing in the bar (inset). The data is shown for a bar defined by identical 
hard walls with $W=1000 l$, $\nu=1$ and $l_0=1.9 l$. For a typical QHE sample under these 
conditions $l\approx 10^{-6}{\rm cm}$ and $e\omega_c\approx 5\mu{\rm A}$. The dotted line 
depicts $\Gamma(k)$ in units of $\omega_c$ assuming a QHE plateau with $\sigma_{xx}/
\sigma_{xy}=5\cdot 10^{-6}$.}
\label{pumpgraph}
\end{figure}

\begin{figure}
\caption{The threshold amplitude of the oscillating current as a function of its frequency. 
The results are presented for various values of constant current through the bar. 
The conditions are the same as in Fig. 1.}
\label{thgraph}
\end{figure}


\begin{references}
\bibitem{Lvov94} V. S. L'vov, {\it Wave Turbulence Under Parametric Excitation} 
(Springer-Verlag, 1994).
\bibitem{Halperin82} B. I. Halperin, Phys. Rev. B {\bf 25}, 2185 (1982).
\bibitem{Wen91a} X. G. Wen, Phys. Rev. B {\bf 43}, 11025 (1991).
\bibitem{Allen83} S. J. Allen, H. L. Stormer and J. C. M. Hwang, Phys. Rev. B 
{\bf 28}, 4875 (1983).
\bibitem{Govorkov86} S. A. Govorkov, M. I. Reznikov, A. P. Senichkin and V. I. 
Talyanskii, JETP Lett. {\bf 44}, 487 (1986).
\bibitem{Volkov86} V. A. Volkov {\it et al.}, JETP Lett. {\bf 44}, 655 (1986).
\bibitem{Govorkov87} S. A. Govorkov {\it et al.}, JETP Lett. {\bf 45}, 316 (1987).
\bibitem{Ashoori92} R. C. Ashoori {\it et al.}, Phys. Rev. B {\bf 45}, 3894 (1992).
\bibitem{Talyanskii92} V. I. Talyanskii {\it et al.}, Phys. Rev. B {\bf 46}, 
12427 (1992).
\bibitem{Volkov88} V. A. Volkov and S. A. Mikhailov, Sov. Phys. JETP {\bf 67}, 
1639 (1988).
\bibitem{Wen91b} X. G. Wen, Phys. Rev. B {\bf 44}, 5708 (1991).
\bibitem{Oreg95} Y. Oreg and A. M. Finkel'stein, Phys. Rev. Lett. {\bf 74}, 3668 (1995).
\bibitem{Orgad95} D. Orgad and S. Levit, cond-mat/9507026 .
\bibitem{Wen92} X. G. Wen, Int. J. Mod. Phys. B {\bf 6}, 1711 (1992).
\bibitem{Stone92} {\it Quantum Hall Effect}, ed. M. Stone (World Scientific, 1992), 
Chap. 5 and references therein.
\bibitem{Prange90} {\it The Quantum Hall Effect}, ed. E. Prange and M. Girvin 
(Springer-Verlag, 1990), pp 51,197 and references therein.
\end{references}
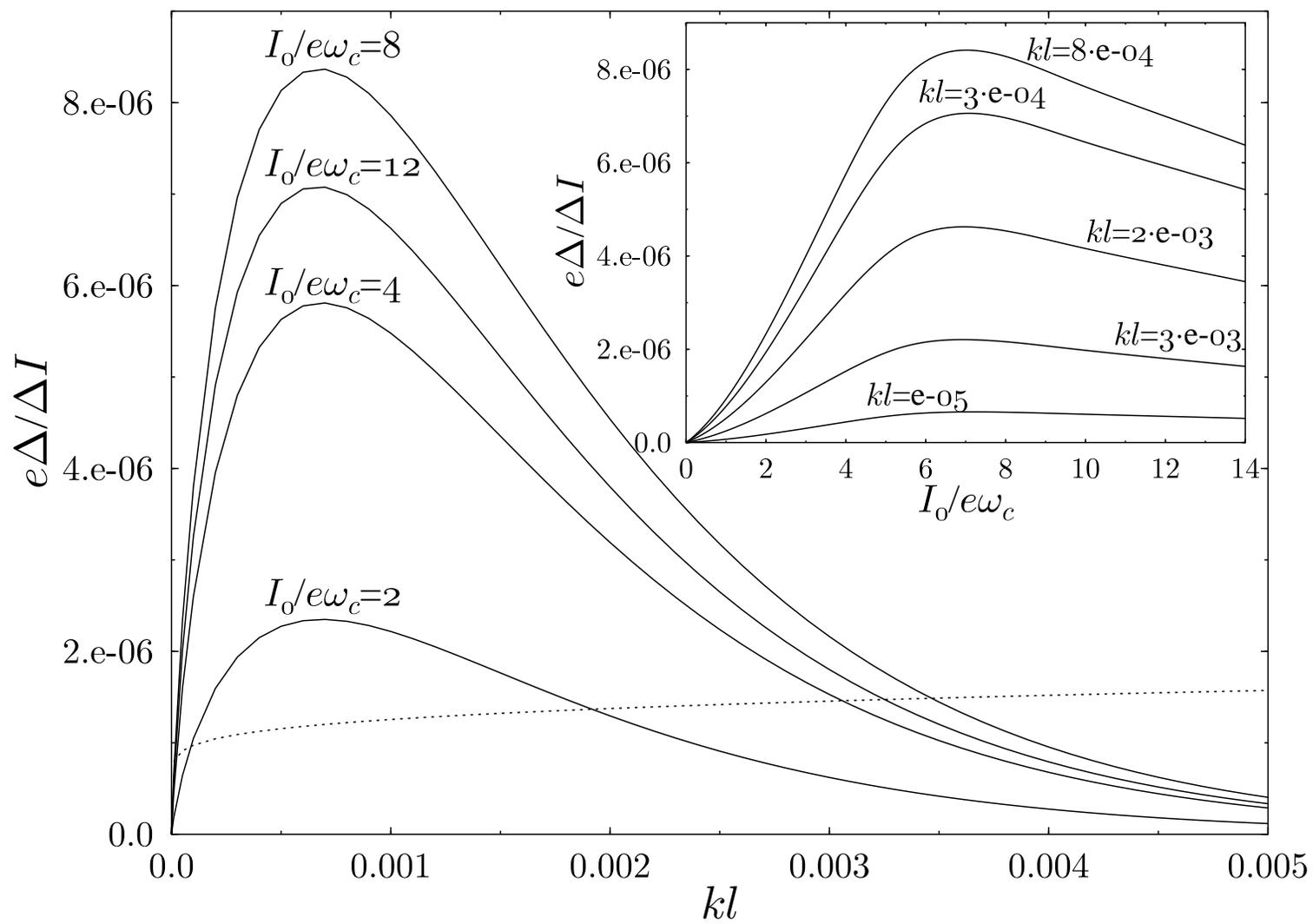
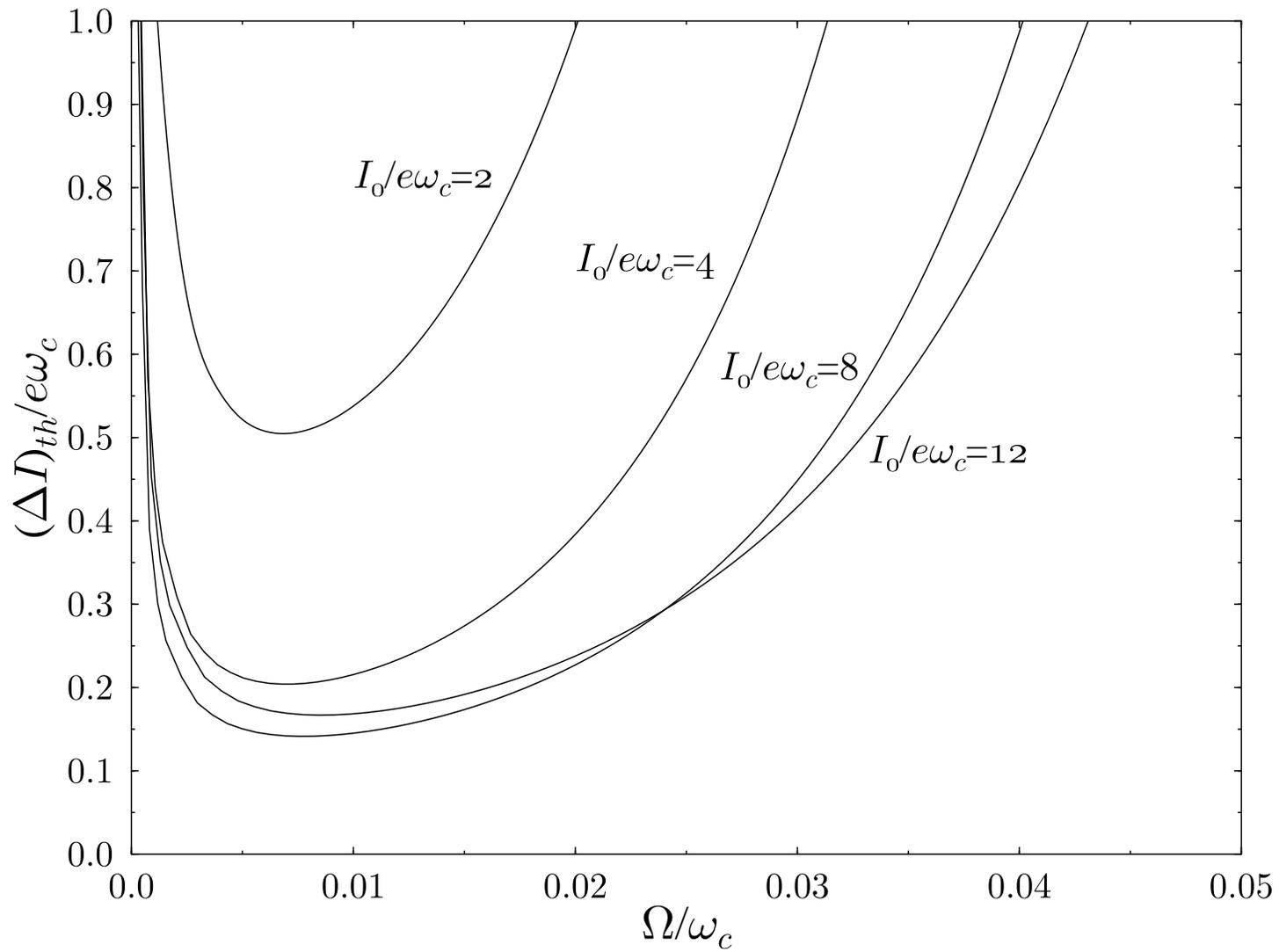
\end{document}